\documentstyle[12pt]{article}
\input epsf

\begin{document}
\baselineskip=15pt
\newcommand{\x}{{\bf x}}
\newcommand{\y}{{\bf y}}
\newcommand{\z}{{\bf z}}
\newcommand{\bp}{{\bf p}}
\newcommand{\A}{{\bf A}}
\newcommand{\B}{{\bf B}}
\newcommand{\p}{\varphi}
\newcommand{\del}{\nabla}
\newcommand{\be}{\begin{equation}}
\newcommand{\ee}{\end{equation}}
\newcommand{\bq}{\begin{eqnarray}}
\newcommand{\eq}{\end{eqnarray}}
\newcommand{\ba}{\begin{eqnarray}}
\newcommand{\ea}{\end{eqnarray}}
\def\r{\nonumber\cr}
\def\hf{\textstyle{1\over2}}
\def\qr{\textstyle{1\over4}}
\def\Sc{Schr\"odinger\,}
\def\sc{Schr\"odinger\,}
\def\'{^\prime}
\def\>{\rangle}
\def\<{\langle}
\def\-{\rightarrow}
\def\dbd{\partial\over\partial}
\def\tr{{\rm tr}}
\def\hg{{\hat g}}
\def\ca{{\cal A}}
\def\pd{\partial}
\def\dl{\delta}

\begin{titlepage}
\vskip1in
\begin{center}
{\large Boundary fermion currents and subleading order chiral
anomaly in the AdS/CFT correspondence.}
\end{center}
\vskip1in
\begin{center}
{\large David Nolland}

\vskip20pt

Department of Mathematical Sciences

University of Liverpool

Liverpool, L69 3BX, England

{\it nolland@liv.ac.uk}

\end{center}
\vskip1in
\begin{abstract}
We construct a wave-functional whose argument couples to boundary
fermion currents in the AdS/CFT correspondence. Using this we
calculate the contributions from bulk fermions to the chiral
anomaly that give the subleading order term in the exact
$N$-dependence of the chiral anomaly of ${\cal N}=4$ SYM. The
result agrees with the calculation of Bilal \& Chu. \noindent

\end{abstract}

\end{titlepage}

The AdS/CFT correspondence is typically studied in the large-$N$
limit, corresponding to the planar limit of the boundary gauge
theory and the classical limit of the bulk supergravity/string
theory. This limit allows us to calculate many interesting
quantities on both sides of the correspondence to leading order in
the large-$N$ expansion. Matching the results then gives us a
dictionary that allows us to translate gauge theory effects into
the language of holographic supergravity.

Although the leading order in $N$ is extremely illuminating by
itself, when we go beyond this approximation to consider quantum
loops in AdS and subleading order effects in the gauge theory, the
dictionary becomes much more subtle. This becomes clear when we
consider the holographic origin of effects such as the Weyl and
R-symmetry anomalies \cite{weyl,r}. At leading order these receive
contributions only from the classical graviton and Chern-Simons
actions, repectively, but at order $1/N^2$ there are one-loop
contributions from all the fields of supergravity.

 One of the first calculations in the AdS/CFT
correspondence of such an order $1/N^2$ effect was the calculation
of Bilal \& Chu of order $1/N^2$ corrections to the R-symmetry
anomaly \cite{chu}. The purpose of this note is to show how the
result of \cite{chu} arises in a wave-functional formalism. Our
method involves the construction of a generating function for
fermion current correlators that may be useful for more general
calculations. In the usual analysis the subleading order
corrections are induced by a Pauli-Villars regularisation that is
introduced to regulate spinor loops in the bulk
\cite{moore,semenoff}. Instead, we use heat-kernel methods to
regulate the generating functional for boundary fermion currents.

In the Type IIB supergravity action the fermions couple to bulk
vector fields that are dual to the boundary R-current. Treating
these gauge vectors as background fields, a one-loop integration
over the fermion fields generates an effective action depending on
the gauge fields. This effective action includes Chern-Simons
terms that contribute to the anomaly. If we interpret the radial
coordinate of AdS as Euclidean "time", the cutoff near the
boundary is a surface on which we can perform canonical
quantisation. According to the standard prescription of AdS/CFT
the Schr\"odinger wave-functional of supergravity fields defined
on this surface is identified with the partition function of the
boundary theory. This wave-functional includes radiatively induced
Chern-Simons terms.

The full action for Type IIB Supergravity compactified on
$AdS_5\times S^5$ has not been determined. However, the
three-point couplings of fermion fields to gauge vectors that we
need are obtained by covariantising the derivatives in the
quadratic action. First, though, let us consider the quadratic
action without couplings to gauge fields. The fermion action is

\be \int d^{d+1}x\sqrt G\bar\psi(\gamma^A D_A-m)\psi, \ee (capital
Roman indices run from 0 to 4) and the bulk metric (incorporating
a curved boundary) is
\begin{equation}
ds^2 = G_{AB}\,dX^A\,dX^B=dr^2 + z^{-2}¥\, e^{\rho}
\hg_{\mu\nu}(x)\, dx^\mu dx^\nu \, ,\quad e^{\rho/2}=
1-C\,z^{2}\,, \quad¥C={l^2 {\hat R} \over
4\,d(d-1)}\,,\label{ads1}
\end{equation}

where $z=\exp(r/l)$ with $l$ a length scale for AdS. We will
interpret $r$ as Euclidean time. The spin-covariant derivative is
defined via the funfbein

\be V^\alpha_0={1\over z}\delta^\alpha_0,\qquad
V^\alpha_\mu={1\over z}e^{\rho/2}\tilde V^\alpha_\mu, \ee

where $\tilde V^\alpha_\mu$ is the vierbein for the boundary
metric (Greek indices run from 1 to 4). Making the change of
variables $\psi=z^2e^{-\rho}\tilde\psi$ causes the volume element
in the path-integral to become the usual flat-space one, and the
kinetic term in the action acquires the usual form. The action can
be written

\be \int d^{d+1}x \bar{\tilde\psi}
\left(\gamma^0\partial_0+ze^{-\rho/2}\gamma^\mu\tilde
D_\mu-m\right)\tilde\psi. \label{act}\ee The $D_\mu$ derivative is
spin-covariant with respect to the boundary metric.

We impose the following boundary conditions on $\tilde\psi$:

\be
Q_+\tilde\psi(\tau,x)=u(x)=Q_+u(x),\qquad\tilde\psi^\dagger(\tau,x)
Q_-=u^\dagger(x)=u^\dagger(x)Q_-,\label{bb} \ee

for $\tau=\exp(r_0/l)$ a small time cutoff on $z$ near the
boundary at $z=0$ and some local projection operators $Q_\pm$. The
remaining projections are represented by functional
differentiation. The partition function takes the form

\be \Psi[u,u^\dagger]=\exp[f+u^\dagger\Gamma u], \label{fw}\ee and
the Schr\"odinger equation that it satisfies can be written

\be {\partial\over\partial r_0}\Psi=-\int d^dx\left(u^\dagger
Q_-+{\delta\over\delta u}Q_+\right)h\left(Q_+
u+Q_-{\delta\over\delta u^\dagger}\right)\Psi, \label{fa}\ee

where $h=\tau e^{-\rho/2}\gamma^0\gamma^\mu\tilde
D_\mu-\gamma^0m$. Note that from the four-dimensional point of
view $\gamma^0$ is what we usually call $\gamma^5$. For the moment
we choose $Q_\pm$ to be $Q_\pm=\hf(1\pm \gamma^4)$. Inserting this
into (\ref{fa}) and (\ref{fw}) gives

\be \dot\Gamma=\tau e^{-\rho/2}D_4-m-2\tau
e^{-\rho/2}\gamma^0\gamma^iD_i\Gamma+\Gamma^2(\tau
e^{-\rho/2}D_4+m)\label{feq}\ee

while the free energy $f$ satisfies

\be \dot f={\rm
Tr}\left(Q_+e^{-\rho/2}\gamma^0\gamma^iD_i+Q_-\Gamma(\tau
e^{-\rho/2}D_4+m)\right).\label{ffe} \ee

Here the index $i$ runs from $1$ to $3$. So that we can regulate
the free energy with a heat-kernel, we expand $\Gamma$ in terms of
the positive-definite operator $(\gamma^\mu\tilde D_\mu)^2$ (this
is possible because of the underlying euclidean invariance):

\be \Gamma=\sum_{n=0}^\infty
\gamma^0(b_n(r_0)+c_n(r_0)\gamma^\mu\tilde D_\mu)(\gamma^\mu\tilde
D_\mu)^{2n}. \ee

Inserting into (\ref{feq}) gives a difference equation for the
coefficients $b_n$ and $c_n$ with a unique solution. So, for
example, $b_0=-1$ and all other coefficients vanish as
$r_0\--\infty$. To regulate (\ref{ffe}) we use a Seeley-de Witt
expansion of the heat-kernel:

\be (\gamma\cdot\tilde D)^{2n}\sim\left(-{\partial\over\partial
s}\right)^{n}\exp\left(-s(\gamma\cdot\tilde D)^2\right), \ee

\be\exp \left(-s(\gamma\cdot\tilde D)^2\right) =\int d^4
x\,{\sqrt\hg}{1\over 16\pi^2s^2}\left(1+s\,a_1(x)
+s^2\,a_2(x)+s^3\,a_3(x)+..\right)\label{sdw} \ee

The contribution to (\ref{ffe}) proportional to the $a_2$
coefficient of $(\gamma\cdot\tilde D)^2$ is finite as $s\-0$ and
$r_0\--\infty$ and as discussed in \cite{us,us2} it determines the
Weyl anomaly. This gives the same result as the calculation in
\cite{us,us2}, as it should for our choice of boundary conditions,
according to the analysis of \cite{fbcs}. The above discussion
extends trivially to boundary conditions given by \be
Q_+=\hf(1\pm\gamma^\mu).\label{fc}\ee

We wish to study the effect of a chiral (R-symmetry)
transformation of the fermion fields. To make the relation to
fermion currents more explicit, we construct a wave-functional
that corresponds to a generating functional of R-current
correlators. The partition function (\ref{fw}) can be expressed as
a path-integral over the upper half-plane cut off at $z=\tau$ so
that

\be \Psi[u^\dagger,u]=\int D\psi^\dagger D\psi e^{-S+\int
d^4x(u^\dagger\psi-\psi^\dagger u)},\ee where $S$ is the action
(\ref{act}), and the boundary conditions are given by (\ref{bb}).
Now consider

\be \Phi[\zeta]=\int Du^\dagger Du\Psi[\zeta u^\dagger,u]e^{\int
d^4x u^\dagger\gamma^5 u}=\int D\psi^\dagger D\psi e^{-S+\int
d^4x\zeta\psi^\dagger Q_+ \gamma^5\psi}.\ee According to the
standard AdS/CFT prescription this is the generating functional
for correlators of the fermion current $\psi^\dagger
Q_+\gamma^5\psi$. So we find

\bq \langle\psi^\dagger
Q_+\gamma^5\psi\rangle&=&\left.{\delta\over\delta\zeta}\det\,^{1/2}(\zeta\Gamma+\gamma^5)\right|_{\zeta=0}\nonumber\\
&=&\hf\tr\left.\left({\Gamma\over\gamma^5+\zeta\Gamma}\right)\right|_{\zeta=0}=
\hf\tr(Q_+\gamma^5\Gamma)\eq

Since this holds for all boundary conditions of the form
(\ref{fc}) we conclude that

\be \langle\bar\psi
\gamma^\mu\psi\rangle=\hf\tr(\gamma^\mu\gamma^5\Gamma)\ee

This vanishes (as we expect) even when couplings to gauge fields
are included in the heat-kernel regularisation of the trace.

The chiral anomaly can be obtained by considering the current
$\langle\psi^\dagger\psi\rangle$ that can be obtained in our
formalism as follows.

At the cost of a spurious enlargement of the Hilbert space, we can
represent fermion operators on the boundary by \cite{fj}

\be \psi^\dagger\sim{1\over
\sqrt2}\left(v^\dagger+{\delta\over\delta v}\right),\qquad
\psi\sim{1\over \sqrt2}\left(v+{\delta\over\delta
v^\dagger}\right). \label{fj}\ee

When we construct solutions of the Schr\"odinger equation using
the representation (\ref{fj}) they take the form

\be \Psi[v^\dagger,v]=\exp[f+2v^\dagger(Q_+-Q_-+Q_-\Gamma Q_+)v],
\ee with $\Gamma$ the same kernel as before, so it may seem as
though not much has changed, but the difference is that
$v^\dagger$ and $v$ are now unconstrained, and we can express

\be\langle \psi^\dagger\psi\rangle=\Phi_0[\zeta]|_{\zeta=0}\ee

\be \Phi_0[\zeta]=\int Dv^\dagger Dv\Psi[\zeta v^\dagger,v]e^{\int
d^4x 2v^\dagger v}.\label{fwf}\ee

Using the same boundary conditions as before, this gives

\be \langle \psi^\dagger\psi\rangle=\hf\tr(Q_+-Q_-+Q_-\Gamma
Q_+)\sim\hf\tr(Q_+-Q_--2Q_-\gamma^5)=\hf\tr\gamma^5.
\label{ffa}\ee It is useful to note that we get the same result if
we use the boundary conditions $Q_\pm=\hf(1\pm\gamma^5)$ and the
corresponding solution found in \cite{us2}, since then $\Gamma$
vanishes as $\tau\to0$. The trace should be regulated with the
heat-kernel expansion (\ref{sdw}) of the operator
$(\gamma\cdot\tilde D)^2$ (covariantised with respect to the gauge
fields as well as the boundary metric). Turning on the background
gauge fields is necessary to give a non-zero result, since there
is no gravitational anomaly.

It is easy to check that the contribution of fermion fields to the
anomaly given by (\ref{ffa}) agrees with the calculation of
\cite{chu}, and therefore summing over the full Kaluza-Klein
spectrum gives the exact N-dependence of the boundary chiral
anomaly.

In conclusion, the subleading order part of the exact N-dependence
of the boundary chiral anomaly coming from spinor loops in the
bulk can be read off from the heat-kernel regularisation of the
wave-functional (\ref{fwf}). This is different from the usual
method for obtaining the anomaly. The formalism may give a useful
way of calculating more general correlators of fermion currents in
the AdS/CFT correspondence.


\begin{thebibliography}{88}

\bibitem{us}
P. Mansfield and D. Nolland, JHEP 9907 (1999), 028.

\bibitem{us2} P. Mansfield, D. Nolland and T. Ueno, JHEP 0401 (2004), 013.

\bibitem{fbcs} D. Nolland, Phys. Lett. B584 (2004), 200.

\bibitem{weyl}M. Henningson and K. Skenderis JHEP 9807 (1998),
023.

\bibitem{r}D.Z. Freedman, S.D. Mathur, A. Matusis and L. Rastelli,
Nucl.Phys. B546 (1999), 96.

\bibitem{Kim}
H.J.~Kim, L.J.~Romans and P.~van Nieuwenhuizen, Phys. Rev. D32
(1985), 389.

\bibitem{warner} D.Z. Freedman, S.S. Gubser, K. Pilch, N.P. Warner,
Adv.Theor.Math.Phys. 3 (1999) 363-417.

\bibitem{chu}A. Bilal and C-S. Chu, Nucl. Phys. B562 (1999), 181.

\bibitem{semenoff}A.J. Niemi and G.W. Semenoff, Phys. Rev. Lett.
51 (1983), 2077.

\bibitem{moore}L. Alvarez-Gaume, S. Della Pietra and G. Moore,
Ann. Phys. 163 (1985), 288.

\bibitem{fj}R. Floreanini and R. Jackiw, Phys. Rev. D37 (1988),
2206.



\end{thebibliography}
\end{document}